\documentstyle[prb,aps,epsf,twocolumn]{revtex}

\tighten

\begin{document}

\title{\bf Event-based relaxation of continuous disordered systems}
\author{G. T. Barkema}

\address{Institute for Advanced Study, Olden Lane, Princeton NJ 08540, USA}

\author{Normand Mousseau}

\address{D{\'e}partement de Physique et Groupe de recherche en physique
et technologie des couches minces (GCM), 
Universit{\'e} de Montr{\'e}al, Montr{\'e}al, Qu{\'e}bec, Canada, H3C 3J7}

\author{
A computational approach is presented to obtain energy-minimized structures
in glassy materials. This approach, the activation-relaxation technique
(ART), achieves its efficiency by focusing on significant changes in the
microscopic structure (events). The application of ART is illustrated with
two examples: the structure of amorphous silicon, and the structure of
Ni$_{80}$P$_{20}$, a metallic glass.
}

\maketitle

Amorphous and glassy materials, from silica glasses to polymers and
proteins, are of considerable fundamental and technological
interest. Progress in the understanding of these materials has been
impaired, however, by the very slow dynamics and relaxation they
display. This has been especially true for computational studies
where machine limitations remain important. In the last years,
significant advances have been made in techniques to study the
dynamics of disordered systems in a discrete space, like spin glasses
and reaction-diffusion models.  Improvements like the Swendsen and
Wang cluster algorithm \cite{cluster} and the event-based method of
Voter \cite{voter} and Barkema {\it et al.}  \cite{red} have increased
the accessible time-scale for these problems by many orders of
magnitude.  In contrast, no such development has been seen for
disordered systems like glassy materials, polymers or proteins. In
these systems, what constitutes a change in configuration ---a move---
is generally ill-defined due to the continuous space to which the
atoms have access: coordination, state, and defects fluctuate widely,
and cannot be characterized uniquely. The generic approach for
studying these continuous disordered systems is molecular dynamics
(MD), which is not sensitive to these problems of definition. However,
time steps in MD simulations are set by the phonon period of 50 to 100
femtoseconds, and therefore MD simulations generally cannot reach the
relaxation time-scale at which many interesting phenomena happen.

Rather than following the irrelevant details of the molecular motion
as atoms vibrate back and forth about their average position, we
propose in this Letter a method which focuses on significant changes
in the microscopic structure (events), i.e., those which imply
crossing the barriers that impair relaxation.  For disordered systems,
which show slow evolution, this event-based technique therefore allows
the study of relaxation on the system's own time-scale, not that of
the phonons.  Of course, once events are defined, it is possible to
either relax the network statically to a global minimum or to follow
its long-time dynamics.

The activation--relaxation technique (ART) allows the system to evolve
following well--defined paths between local energy minima. Starting in
a local minimum, it proceeds in two steps:
\begin{description}
\item[Activation:] The configuration is moved to a local saddle--point.
\item[Relaxation:] The system is pushed over the saddle--point and relaxed
to a new minimum.
\end{description}
By restricting its search of paths to those going through
saddle--points, ART focuses on the minuscule part of the configuration
space which is physically accessible and therefore results in an
efficiency which cannot be achieved by more conventional Monte Carlo
moves.

We first discuss the implementation of ART in more detail.
We then illustrate the application of ART by considering two test--cases
for glassy materials: amorphous silicon and Ni$_{80}$P$_{20}$, a metallic 
glass. We conclude with a discussion of other applications for ART.

\begin{figure}
\epsfxsize=8cm
\epsfbox{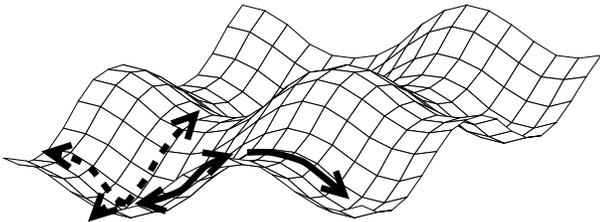}
\caption{ Illustration of the activation and relaxation steps in
ART for a two--dimensional energy landscape. The bottom of the wells
represent local minima. From each minimum, the system can reach four
saddle--points. The four arrows pointing up in the front right handside
well indicate the possible directions of $\vec{G}$ as defined in Eq. 1.
Similarly, the down-pointing arrows from the saddle-point in the center
indicate the directions of the force $\vec{F}$ toward a minimum energy. The
full-line arrows show the path for a single event.}
\label{saddle}
\end{figure}

{\it Algorithm---} The underlying concept of ART is that it is the
material itself, as defined by its total energy surface, which should
determine events, not an external rule.  This idea is realized through
the following steps: We start with a configuration relaxed in a local
energy minimum using a force $\vec{F}$ derived from an appropriate
interaction potential (empirical or {\it ab initio}). At the start of
the activation, a single atom is slightly displaced along a random
direction to create a non-zero term in the force. The configuration is
then moved to a nearby saddle--point by iterative application of a
redefined force:
\begin{equation}
\vec{G}=\vec{F}-(1+\alpha) (\vec{F}\cdot \Delta \hat{X}) \Delta \hat{X}.
\label{eq:1}
\end{equation}
Here, $\vec{F}$ is the $3N$-dimensional force vector from the
interaction potential, $\hat{X}$ is a $3N$-dimensional unit vector
pointing from the last local minimum to the current position of the
configuration and $\alpha$ is a positive number. The new force
$\vec{G}$ is thus opposite in sign to $\vec{F}$ in the direction {\it
parallel} to $\hat{X}$, and equal to $\vec{F}$ in any direction {\it
perpendicular} to $\hat{X}$. At each iteration, $\vec{G}$ is evaluated
and the atoms are moved according to it, therefore forcing the system
to follow a valley up--hill in the energy landscape towards a
saddle-point (see Fig. \ref{saddle}.) At the saddle--point, both
$\vec{G}$ and $\vec{F}$ are zero and the iterative process stops: the
activation is complete. We then proceed with the relaxation by moving
the system along $\vec{F}$ into a new local minimum.

Eq. \ref{eq:1} involves all atoms of the system; it is therefore the
whole configuration which is taken through a saddle--point to a new
minimum, not a fixed or pre-defined set of atoms. An event can be
characterized {\it a posteriori} by counting how many atoms have been
displaced significantly (by, say, more than 0.01\AA.)  In this way, we
have found events involving from one to a few hundreds atoms,
indicating that a wide range of excitations can indeed be sampled.

In effect, ART limits the configuration space as
seen by the system. For an $N$-particle system, this space is a
$3N$-dimensional continuum; however, most of it is energetically
highly unfavorable, and thus never visited if the system is at low
temperature. In fact, the configuration is almost always located in or
very near a local energy minimum. Thus the system is effectively
confined to ``dots'' in the $3N$--dimensional configuration space,
corresponding to the local energy minima. Given enough time, however,
the configuration will hop from one such local minimum to another
nearby. To take this hopping into account, we connect the isolated
dots by paths going through local saddle--points, which are the most
likely, thus creating a whole network of minima. This reduction of a
$3N$-dimensional space to an effective configuration space consisting
of the local energy minima and paths between them makes possible the
efficient relaxation obtained with ART. This does not mean that we
ignore the phonons altogether. They remain indirectly present in
ART, as they provide the energy for going through the saddle--points,
i.e., jumping the barriers. This is sufficient since the relaxation and
dynamics of glasses and amorphous materials are orders of magnitude
slower than the typical phonon period or, in other words, the
energy barriers in these systems are much larger than the temperature
at which these materials are studied.

On a more technical side, we have chosen the Levenberg--Marquardt
minimization technique both for activation to the saddle--point and
relaxation away from it. To perform global minimization, we used
simulated annealing.  The acceptance ratio is given by
$P_{accept}={\mbox Min} \left[ 1,\exp(-\Delta E/kT) \right]$, where
$T$ is a fictive temperature.  It is also necessary to avoid
saddle--points found in the very close vicinity of the minimum and
which are an artifact of the static simulation. In an MD simulation,
these low saddle--points would be washed out by the constant vibration
of the lattice. In order to shun them, we introduced a repulsive
harmonic potential which forces the system away from these unimportant
saddle--points, $E = A ( |\Delta \vec{X}|-X_c)^2$ with a cut-off at
$X_c$.  $|\Delta \vec{X}|$ represents the displacement of the whole
configuration, and $X_c$ is typically 0.5\AA.

{\it Amorphous silicon---} Amorphous silicon ({\it a}-Si) represents a
good test-case for the application of ART to covalent amorphous
materials. Many techniques have been used to study this material,
including the static bond-switching algorithm of Wooten, Winer and
Weaire (WWW) \cite{www}, which proved very efficient in spite of the
unphysical events it proposes, and MD with a wide choice of
interactions \cite{md}. The MD models all suffer from the presence of
a large number of defects, most of which, moreover, being floating
bonds, i.e. overcoordinated atoms, which are not seen
experimentally. This is particularly so when MD is carried out with
empirical potentials. Even the best of those, developed by Stillinger
and Weber (SW)\cite{sw}, generally gives a structure with a
coordination of $4.12$--$4.30$. Tight-binding \cite{tb} and {\it
ab-initio} MD \cite{abinitio} also show too many defects but the
coordination is normally lower; much longer relaxation with larger
unit cells should result in a good agreement with experiment. The
situation is not so clear in the case of empirical potentials, where
doubts remain as to whether the problem comes from the potential
itself or from too short simulations\cite{question}. ART is well
suited to address this question, since it does not suffer from the
slow dynamics inherent to MD.

\begin{figure}
\epsfxsize=8cm
\epsfbox{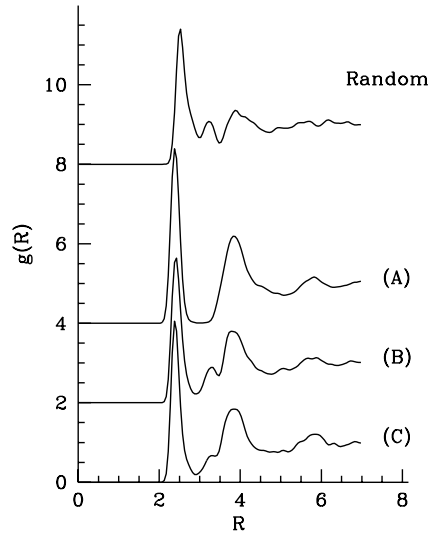}
\caption{Radial distribution function for a randomly packed system (top
curve), and after three different energy minimizations:
(A) after relaxation with a modified SW potential with ART
(B) after relaxation with the SW potential with ART 
(C) relaxation with the SW potential with ART, starting from (A)
\label{fig2}}
\end{figure}

First, we show that ART can reproduce the structure found by MD using
the SW potential. Starting from a 216--atom random closed packed
configuration, with a density corresponding to the one of crystalline
silicon (see top curve in Fig. \ref{fig2}), we applied ART until the
system came to a stable energy. The radial distribution function (RDF)
for this configuration is labeled (B) in Fig. \ref{fig2}. The final
energy, $-4.12$ eV/atom, is as low as what is obtained with MD
\cite{kluge} while the coordination of $4.17$ is at the lower end of
the spectrum.  Although in good agreement with MD simulations, the
overcoordination as well as the shoulder on the left side of the
second--neighbor peak in the RDF (curve B in figure \ref{fig2}) show
that the configuration retains a strong liquid--like structure which
is not present experimentally.

To show that this failure to reach the experimental structure is not
due to ART itself, we performed a second simulation with an {\it ad
hoc} modification to the SW potential: the three--body contribution to
the total interaction energy was doubled. The effect of this
modification is to increase the cost of over--coordination, helping
the system to move away from the liquid--like phase. In order to test
the scaling of the method with size, we applied ART, with this
modified potential, to the 216--atom random closed packed
configuration discussed in the previous paragraph and to a 1000-atom
one.  Both networks necessitated about 5 ART--events per atom to reach
a stable point showing that the activation--relaxation technique is
not slowed down with increasing size.  Further simulation did not
change the properties of those networks while structural properties
for the two lattices were essentially the same. The RDF for the
1000--atom configuration is plotted as curve (A) in
Fig. \ref{fig2}. This quantity, as well as an almost perfect
coordination of 3.97 and a rms angular deviation of 9.97 degrees, are
all in excellent agreement with experiment.

Further insight into the cause of the discrepancies between
experiments and SW--based MD simulations was obtained by a third
minimization. Starting this time from the 216-atom configuration
obtained with modified SW, we relaxed it using the conventional SW
parameters. The result, curve C in Fig. \ref{fig2}, shows a liquid-like
structure very similar to the one obtained in (B) with an equivalent
energy per atom of $-4.12$ eV: ART converged towards the same point
starting from completely different initial configurations.

These results indicate clearly that ART can move through the
configuration space towards a minimum in spite of large barriers.  We
could recover the characteristic structure for the SW from two
completely different initial configurations. Moreover, with a modified
potential, it was possible to obtain an amorphous structure in good
agreement with experiment. As a result of our ART simulations, we showed
that the structural
differences between MD with empirical potentials and experiments are
caused, in part, by defective potentials, as was proposed in
Ref. \cite{question}. We expect that ART simulations with
semi-empirical tight--binding interactions or {\it ab initio}
potentials will also produce a network with a structure close to that
observed experimentally.

A final point about these simulations. Since the density of the final
configuration does not have to be the same as for crystalline silicon
({\it c}-Si), we also let the volume relax as one additional degree of
freedom, i.e. the cell remains cubic but fluctuates in size. In the
the first simulation, with SW parameters, this freedom resulted in a
density about 4.5 percent larger than that of {\it c}--Si while the
stronger three--body term in the second simulation leads to a density
which is about 5 percent lower. The final density of configuration in
the last simulation is essentially equal to the crystalline one. This
supports the usual view that the energy landscape of disordered
systems has multiple minima that are very close in energy.

{\it Glassy Ni$_{80}$P$_{20}$---} As the second test-case of ART, we
considered Ni$_{80}$P$_{20}$, a prototypical metallic glass which has
been the subject of considerable investigation (e.g.
\cite{lewis,kob}.) The structure of this glass corresponds to the
dense random packing model proposed by Bernal \cite{bernal}.  Its
topology is controlled by local packing, not covalent bonding, and is
thus relevant for high coordination glasses. The more compact
environment leads to a topological rigidity larger than that for
amorphous semiconductors\cite{rigid}, which makes barriers higher.

\begin{figure}
\epsfxsize=8cm
\epsfbox{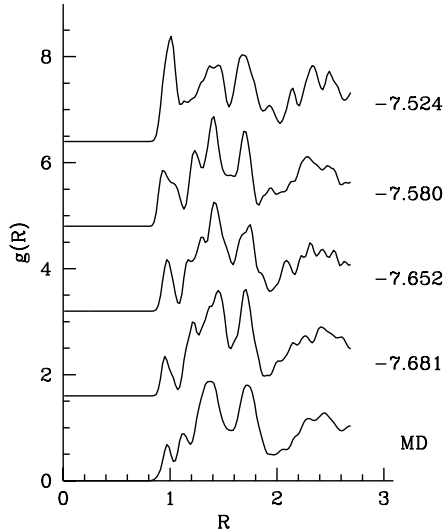}
\caption{Evolution of the radial distribution function of the P-P distances
during the ART simulation. The energy, shown to the right of each curve, is in
Lennard Jones units. The bottom curve is the RDF of a zero-temperature
relaxed MD configuration of 1000 atoms, obtained form Kob and Andersen,
with an energy of -7.667 LJ units.}
\label{fig3}
\end{figure}

We selected a Lennard Jones (LJ) potential with the parameters used by
Kob and Andersen \cite{kob}. This allows for a direct comparison with
their MD simulation that lasted 60 ns. As in the previous section, we
started from a purely random configuration of 250 atoms with the
proper density and evolved it with ART using simulated annealing. The
volume was allowed to adjust. The top four curves in figure \ref{fig3}
show the evolution of the RDF for the P-P distances during the
minimization process. We can see that there is considerable change in
the coordination of P atoms as the total configurational energy decreases.

As discussed recently by Angell \cite{angell}, the lack of low-energy
disordered configurations hinders the study of rapid and slow dynamics
in glasses far away from the glass temperature. The final ART
configuration has an energy of -7.681 LJ units providing a significant
improvement over the excellent MD simulation of Kob and Andersen,
which reaches an energy of -7.667 LJ units when relaxed to zero
temperature (see the bottom curve of figure 3).  From Fig. 3, one can
also see that the number of P-P first neighbors decreases as these
atoms diffuse away from each other in an important rearrangement of
the lattice.  ART can thus be used to obtain a low-energy structure
which can then be studied dynamically either via molecular or
event--based dynamics.

{\it Conclusion---} We have presented a computational approach to the
relaxation of continuous disordered systems that we call
activation--relaxation technique (ART).  ART is not based on any
specific potential or structure and can therefore be applied to a
wide range of systems. In two test-cases, the relaxation of {\it a}-Si
and Ni$_{80}$P$_{20}$ starting from random packed configurations, ART
has provided new insights.  It resolved a long-standing question about
the cause of the liquid--like nature of SW-based MD-relaxed {\it
a}-Si. It also produced Ni$_{80}$P$_{20}$ samples with a significantly
lower energy than MD, showing the way for creating weakly strained
configurations for disordered structures.

For which systems would ART be appropriate? The natural time-scale of ART
is event--based, that of MD is phonon--based. This makes the two techniques
complementary: for a system with short-time dynamics, MD is preferable;
when rare events dominate, ART has to be applied.

In the near future we will combine ART with tight-binding calculations for
the study of {\it a}-Si. This will be extended to other disordered
materials like {\it a}-GaAs and SiO$_2$. Future work will include the
relaxation of polymers and proteins inside and outside solution. We also
want to combine ART with rare--event dynamics \cite{red}, so that dynamical
quantities like self-diffusion can be obtained with the efficiency of ART.

We thank G. Boisvert, J. L. Brebner and L. J. Lewis for critical
reading of this manuscript.
GTB acknowledges support from the DOE under grant DE-FG02-90ER40542 and from
the Monell foundation. 
\bibliographystyle{prsty}

\end{document}